\documentclass[12pt,manuscript]{aastex}
\usepackage{emulateapj5}
\usepackage{psfig,times}

\newcommand{\snks}{{SN~1006}}

\newcommand{\Te}{$kT_{\rm e}$}
\newcommand{\net}{$n_{\rm e}t$}

\newcommand{\spex}{{\it SPEX}}
\newcommand{\xspec}{{\it xspec}}

\newcommand{\xmm}{{\it XMM-Newton}}
\newcommand{\chandra}{{\it Chandra}}

\begin{document}

\title{Slow temperature equilibration behind the shock front of SN 1006}

\shortauthors{J. Vink et al.}

\author{Jacco Vink\altaffilmark{1,2}, J. Martin Laming\altaffilmark{3}, Ming Feng Gu\altaffilmark{2,4}
	Andrew Rasmussen\altaffilmark{1}, Jelle S. Kaastra\altaffilmark{5}}

%\email {jvink@astro.columbia.edu}
\altaffiltext{1}{Columbia Astrophysics Laboratory, Columbia University, MC 5247, 
550 W 120th street, New York, NY 10027, USA;jvink@astro.columbia.edu}
\altaffiltext{2}{Chandra fellow}
\altaffiltext{3}{Naval Research Laboratory, Code 7674L, Washington DC 20375, USA}
\altaffiltext{4}{Center for Space Research, MIT, Cambridge, MA 02139}
\altaffiltext{5}{SRON National Institute for Space Research, 
Sorbonnelaan 2, NL-3584 CA, Utrecht, The Netherlands}

\begin{abstract}
We report on the observation of O~VII Doppler 
line broadening in a compact knot at
the edge of SN 1006 detected with the Reflective Grating
Spectrometer on board \xmm.
The observed line width of $\sigma = 3.4\pm 0.5$~eV 
at a line energy of 574~eV indicates
an oxygen temperature of $kT = 528\pm150$~keV. 
Combined with the observed electron
temperature of $\sim 1.5$~keV the observed broadening is direct evidence 
for temperature non-equilibration in high Mach number shocks, and slow
subsequent equilibration.
The O~VII line emission allows an accurate determination of the ionization state of the plasma,
which is characterized by a relatively high forbidden line contribution, 
indicating $\log(n_{\rm e} t) \simeq 9.2$.
\end{abstract}

\keywords{
shock waves -- 
X-rays: observations individual (SN 1006) -- 
supernova remnants
}

\section{Introduction}
It has been suspected for many years that the high Mach number, 
collisionless shocks of young supernova remnants (SNRs) 
do not produce electron-ion temperature equilibration.
The Rankine-Hugoniot relations require that for very high Mach numbers
the temperature for each particle species $i$:
\begin{equation}%\begin{eqnarray}
%	\frac{\rho_{2,i}}{\rho_{1,i}} = \frac{\gamma +1}{\gamma -1} = 4 \\
	kT_{i} = \frac{2(\gamma-1)}{(\gamma+1)^2} m_i v_s^2  = \frac{3}{16} m_i v_s^2,
	\label{eq-shocks}
\end{equation}%\end{eqnarray}
with $m_i$ the particle mass, 
$v_s$ the shock velocity, and $\gamma$ the ratio of specific heats, 
usually taken to be $5/3$ \citep[e.g.][]{mckee80}.
The heating process in collisionless shocks is not well understood, 
but the Coulomb collision times are too long to provide the required
heating. So other, collective, processes should be responsible for
the heating.
This raises the question whether the heating process leads to temperature
equilibration or not, i.e. is the electron temperature very low compared to the
proton temperature, which, according to (\ref{eq-shocks}), 
should be lower than the oxygen or iron temperature?
If temperatures are not equilibrated at the shock front, 
and subsequent equilibration
proceeds through Coulomb interactions,
full equilibration takes %approximately 
$\sim 10^{12}/n_{\rm e}$~s
\citep[see][]{itoh84}.

A clear hint for non-equilibration 
is the low electron temperature in young SNRs,
which in no object seem to exceed 5~keV,
whereas a typical shock velocity of $4000$~km/s should give rise to
a mean plasma temperature of 19~keV \citep[e.g.][]{mckee80,hughes00b}.
X-ray observations usually allow only the electron temperature to be 
determined from the continuum shape and line ratios.
The ion temperature is difficult to measure, as it does not alter the
continuum shape, and hardly influences the ionization and excitation balance.

Some measurements of temperature non-equilibration based on optical and 
UV spectroscopy have been reported.
The shock velocity and amount of electron-proton 
equilibration can be determined from
the line widths and ratio of H$\alpha$\ and H$\beta$ emission
from non-radiative shocks.
For the northwestern shock front in \snks, which is also
the subject of this letter, \citet{ghavamian02} measured a
shock velocity from the  H$\alpha$\ 
width of $v_s = 2890$~km/s and inferred an electron to proton temperature
ratio $T_{\rm e}/T_{\rm p} < 0.07$.
Together with the measured proper motion \citet{winkler02} were able to
determine a distance to SN 1006 of $2.17\pm0.08$~kpc.
UV spectra obtained with the Hopkins Ultraviolet Telescope
showed broad C~IV, N~V and O~VI lines, indicating temperature
non-equilibration of these elements \citep{raymond95},
and a comparison of these line intensities with He~II allowed a measurement
of $T_{\rm e}/T_{\rm p} < 0.05$\ \citep{laming96}.
X-ray observations of \snks\ indicate a low
value of typically $\log(n_{\rm e}t) \sim 9.5$\ \citep{vink00a,dyer01}, 
implying that
if the equilibration is governed by Coulomb interactions, the plasma
did not have sufficient time to equilibrate.

Here we report on a direct X-ray measurement of the O~VII temperature 
in the northwest of \snks. It confirms with a high statistical confidence 
the slow equilibration of electron and ion temperature, 
but for a position further downstream from the shock
than for the optical and UV measurements.

\begin{figure*}[ht]
\centerline{
	\psfig{figure=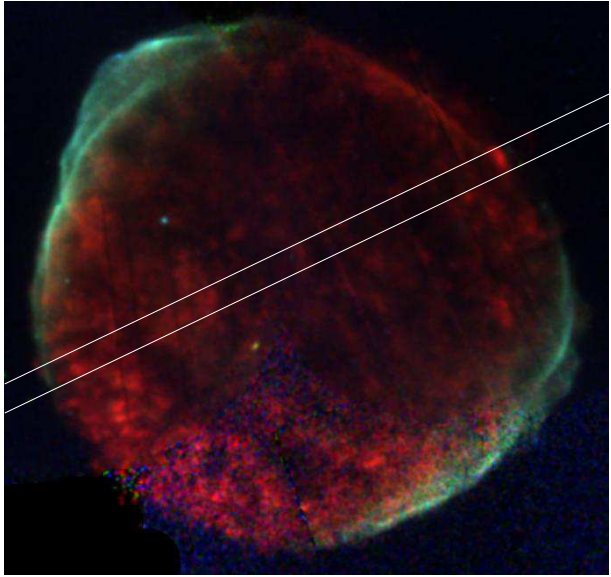,width=8.5cm}
	\psfig{figure=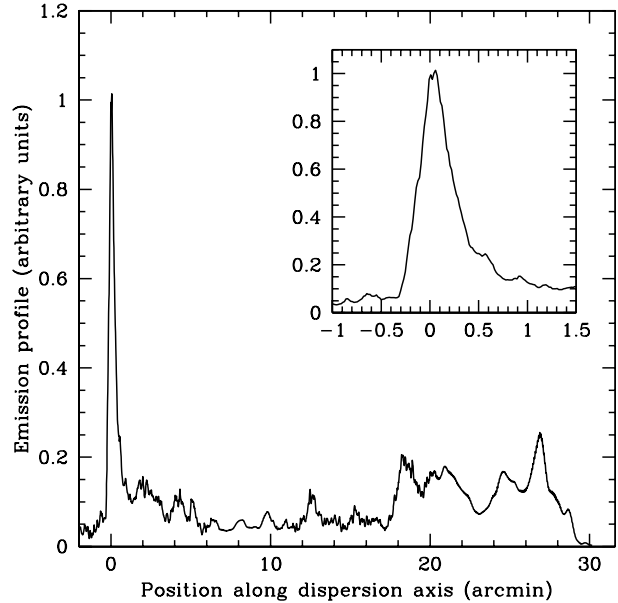,width=8.5cm}
}
 \caption{{\em Left:} 
\xmm\ image of \snks\ with the following color coding:
red, 0.5-0.61 keV (dominated by O VII He$\alpha$); green, 0.75 - 1.6 keV; blue,
1.6 - 7 keV.
The area contributing to the RGS spectra is indicated by the two lines.
{\em Right:} OVII emission profile along the dispersion direction of
the RGS (based on a combination of
EPIC and \chandra\ data). The inset shows the
profile of the bright northwestern knot (\chandra).
\label{fig-prof}}
\end{figure*}

\section{Observations and methods}
We observed \snks\ (G327.6+14.6) with the \xmm\ X-ray observatory  \citep{jansen01}
with the goal of measuring the ionization and equilibration
in \snks.
The two observations are part of the guest observer 
program and were made on August 10 and 11, 2001. One observation 
pointed on the bright knot
in the northwest, the main topic of this {\em letter}, 
the other on the narrow filaments in the east of \snks.
The exposure times varied per instrument, but were close to 60~ks for both observations.
In order to produce Fig.~\ref{fig-prof} we also used additional archival \xmm\ and
\chandra\ data.

Our main result is obtained with the 
Reflective Grating Spectrometer \citep[RGS,][]{denherder01}.
The two RGS instruments, RGS1 and RGS2 cover the
wavelength range of $\sim 6 -  40$~\AA, 
with a first order resolution of $\lambda/\Delta\lambda \sim 300$ at 
20~\AA. Emissions from different orders are separated 
using the intrinsic energy resolution of the CCDs.
Unfortunately one CCD chip on the RGS1 and one on the RGS2 no longer function.
As a result the RGS1 spectrum does not cover the wavelength range encompassing
Ne line emission,
and the RGS2 does not cover the wavelength range around 22~\AA, 
which contains the O~VII He$\alpha$ line emission. 
We limited the extraction region to an 1\arcmin\ wide strip across
the bright northwestern knot.

\bigskip
\vbox{
\centerline{
	\psfig{figure=sn1006nw_epic_brspot.ps,angle=-90,width=8cm}}
\figcaption{EPIC-CCD spectra of the bright knot in the northwest of SN 1006.
The solid line is the best fit single NEI model (see Table~1).
The top spectrum is the EPIC-PN (x5), the bottom spectrum is the combined
EPIC-MOS spectrum. \label{fig-epic}}
}
\bigskip

The extended emission from \snks\ means that the 
standard response matrices had to be convolved with the spatial 
emission profile displayed in Fig.~\ref{fig-prof}, with
an attenuation due to off-axis telescope vignetting.
%This was done with proprietary software. Note, however, 
%that the \xspec\ spectral analysis program \citep{xspec} contains
%a multiplicative model (\rgsxsrc) that produces similar results.
The bright knot has a spatial width of 0.4\arcmin\ (FWHM), which 
gives an apparent
spectral broadening of $\Delta \lambda = 0.05~\AA$ (FWHM), or 
$\lambda/\Delta \lambda = 430$ at 21.6~\AA.
The intrinsic resolution of the RGS is $\Delta \lambda = 0.06$~\AA
\citep{denherder01}. 
%This indicates that Doppler motions as low as $\sim 1000$~km/s 
%can in principle be measured. 
As the knot is compact and 
close to the edge of \snks\ it is unlikely that
any measurable broadening is due to bulk  motions along the line of sight.

\xmm\ also has CCD cameras
behind each of its three mirrors, called the 
European Photon Imaging Camera (EPIC).
We use the spectra extracted from the EPIC data to measure the electron
temperature and abundances. 
All standard data reductions and response matrix calculations were done with SAS 5.3.3.
Background spectra for the EPIC data were taken from a region outside \snks, but 
close to the center of the field.
As RGS background spectra we used archival data of targets with no apparent line emission, such as
gamma ray bursts.

\section{The electron and ion temperatures in the northwest of SN 1006}
The measurement of the non-equilibration of the electron and ion temperatures
requires the measurement of both the ion temperature, here oxygen temperature,
and the electron temperature.

The electron temperature is in this case most accurately measured from the spectral 
continuum shaped observed with the EPIC CCD spectra,
as the low ionization time scales in SN 1006 make the available line ratios
only weakly temperature dependent.
The EPIC spectra from the northwestern knot were fitted with the \spex\
non-equilibration ionization (NEI) code \citep{kaastra96}. As the knot is
relatively compact, temperature and ionization gradients are probably
of minor importance. 
The plasma parameters obtained by fitting the spectra indicate 
$\log(n_{\rm e}t) = 9.4$, and \Te $\sim 1.3-1.7$~keV,
higher than the \Te$\sim$0.7~keV reported by \citet{long03} based 
on \chandra\ data.
Note, however, that the \chandra\ spectra have a lower spectral resolution,
and have currently more calibration problems. 
Moreover, as there are two different
kind of EPIC instruments, we were able to verify the consistency of the 
results (Fig.~\ref{fig-epic}). 
%The spectra from both instruments show continuum emission above 2~keV 
%(Fig.~\ref{fig-epic}).
Nevertheless, the measured \Te\ is substantially lower than 
the $kT_{\rm e} \sim 20$~keV expected for a fully equilibrated plasma,
but still higher than that expected from
Coulomb equilibration alone behind a 3000 km/s shock. Fig.~\ref{fig-eq}
shows the predicted \Te\ against \net\ for current observations of
the SN 1006 knot, as well as the various ion temperatures. Our
fitted \net\ corresponds to plasma shocked 200-300 years ago, which
is predicted to have $kT_{\rm e} = 600$~eV. 
Hence we infer a small degree
of collisionless electron heating (around 5\% of the shock energy, 
i.e. $T_{\rm e}/T_{\rm p} \sim 0.1$, see \citet{ghavamian01}) 
consistent with optical and UV observations \citep{ghavamian02,laming96} 
of SN 1006 and with observations of high Mach number shocks
in other supernova remnants such as Tycho \citep{ghavamian01} 
and SN~1987A \citep{michael02}.

\bigskip
\vbox{
\centerline{\psfig{figure=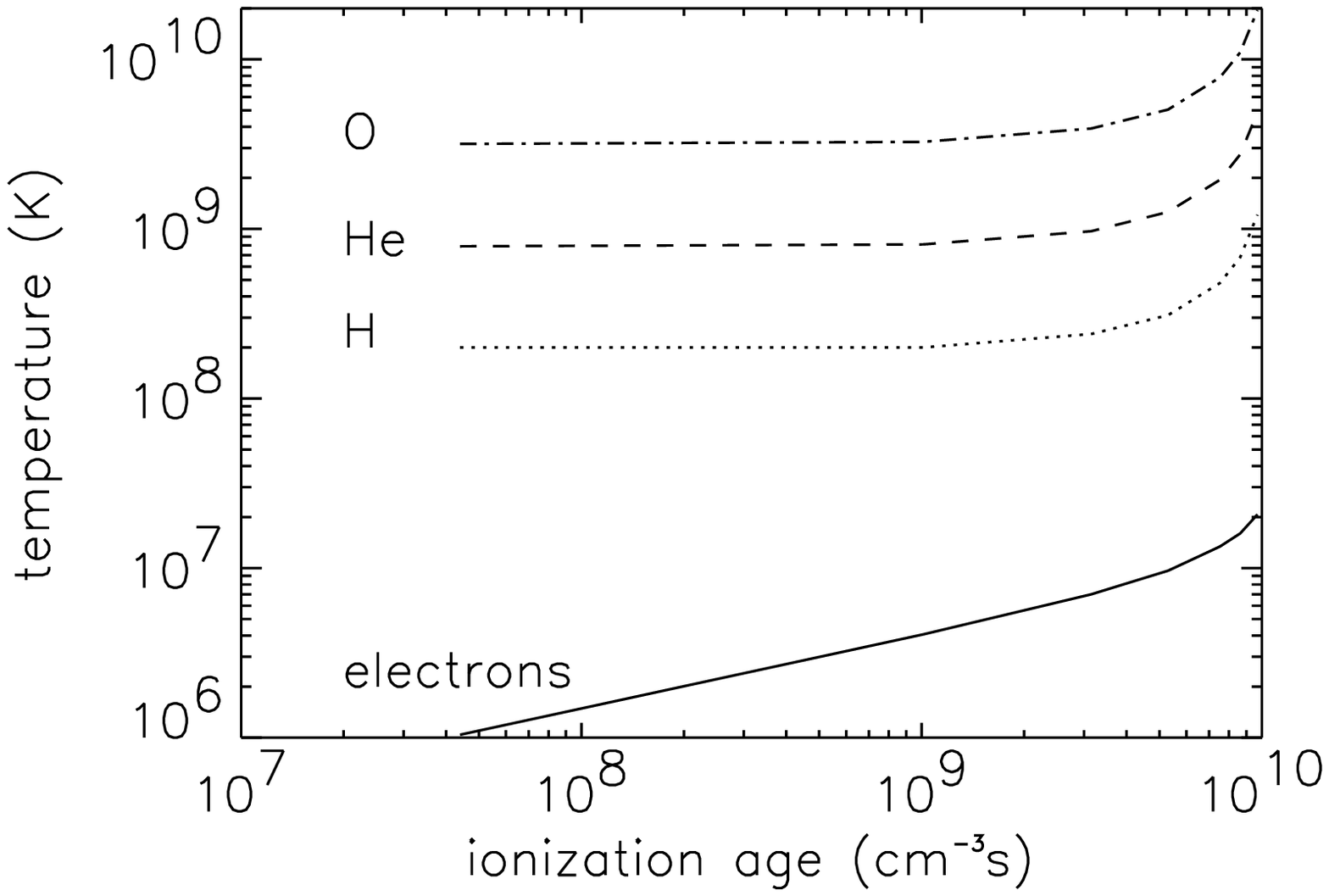,width=8.5cm}}
\figcaption{
Expected relation between \net\ and electron and ion temperature based
on a shock history model for the \snks\ blast wave \citep{truelove99,laming01b}.
\label{fig-eq}}
}
\bigskip

The spectra are dominated by line emission from O~VII, the other line complexes
are, however, not from helium-like stages of Ne, Mg and Si, as identified by
\cite{long03}, but from lower ionization stages. 
For instance the EPIC spectra
show consistently that the Mg line centroid is $1.33\pm0.01$~keV,
whereas the Mg XI line centroid is $1.35$~keV. 
The Si line centroid is
$1.80\pm0.01$~keV, which differs significantly from the
Si XIII centroid of 1.85~keV.
Instead these centroids indicate ionization stages around Mg IX and Si IX.
The Ne centroid, as determined from the RGS2 spectrum, 
indicates $0.900\pm0.004$~keV, consistent with Ne~VII.
These centroids provide clear evidence for extreme NEI conditions for the 
northwestern knot and corroborate the measured \net\ value.

\psfig{figure=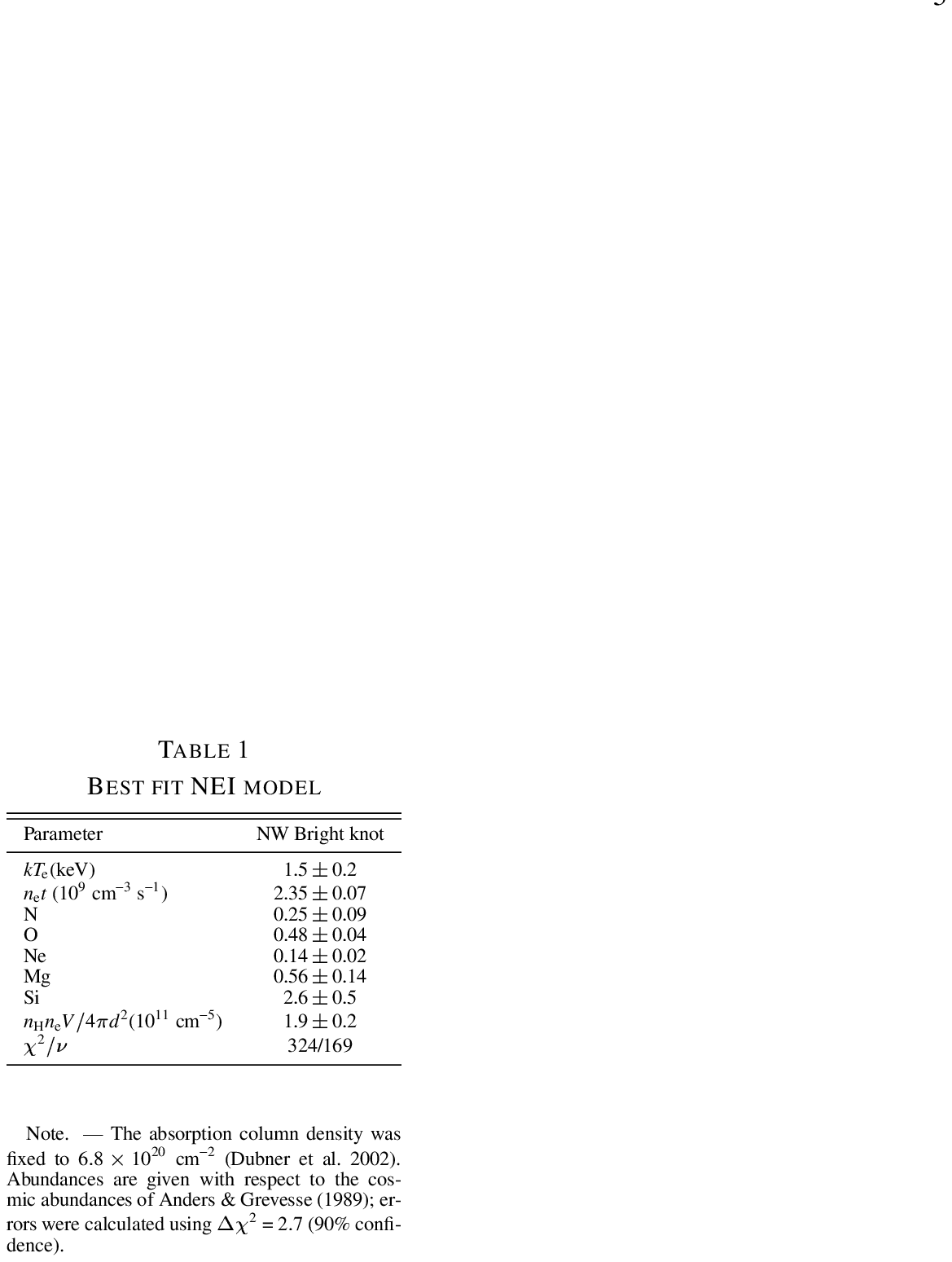}

In agreement with \citet{long03} we find that the knot emission indicates an 
overabundance of Si, but contrary to the \chandra\ data we do find O, Ne and 
%Mg to be somewhat underabundant (Table~\ref{tbl-epic}).
Mg to be somewhat underabundant (Table~1).
This may indicate that instead of shocked ISM the knot is actually ejecta or 
ISM mixed with ejecta. This does not diminish the evidence for temperature 
non-equilibration, but it makes it harder to combine the results reported 
here with the H$\alpha$ shock velocity
measurements.

\begin{figure*}
\centerline{
	\psfig{figure=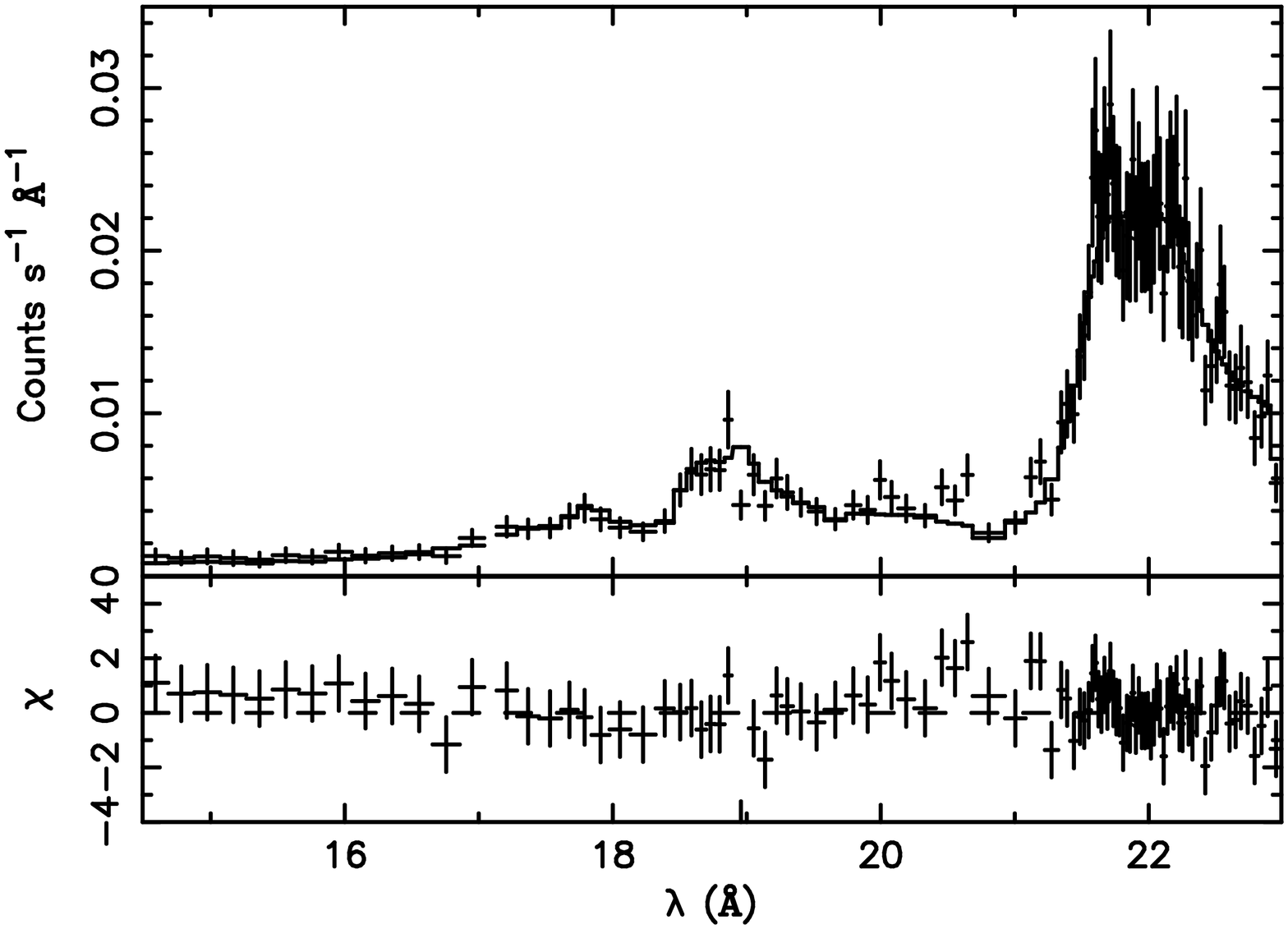,angle=-90,width=8.5cm}
	\psfig{figure=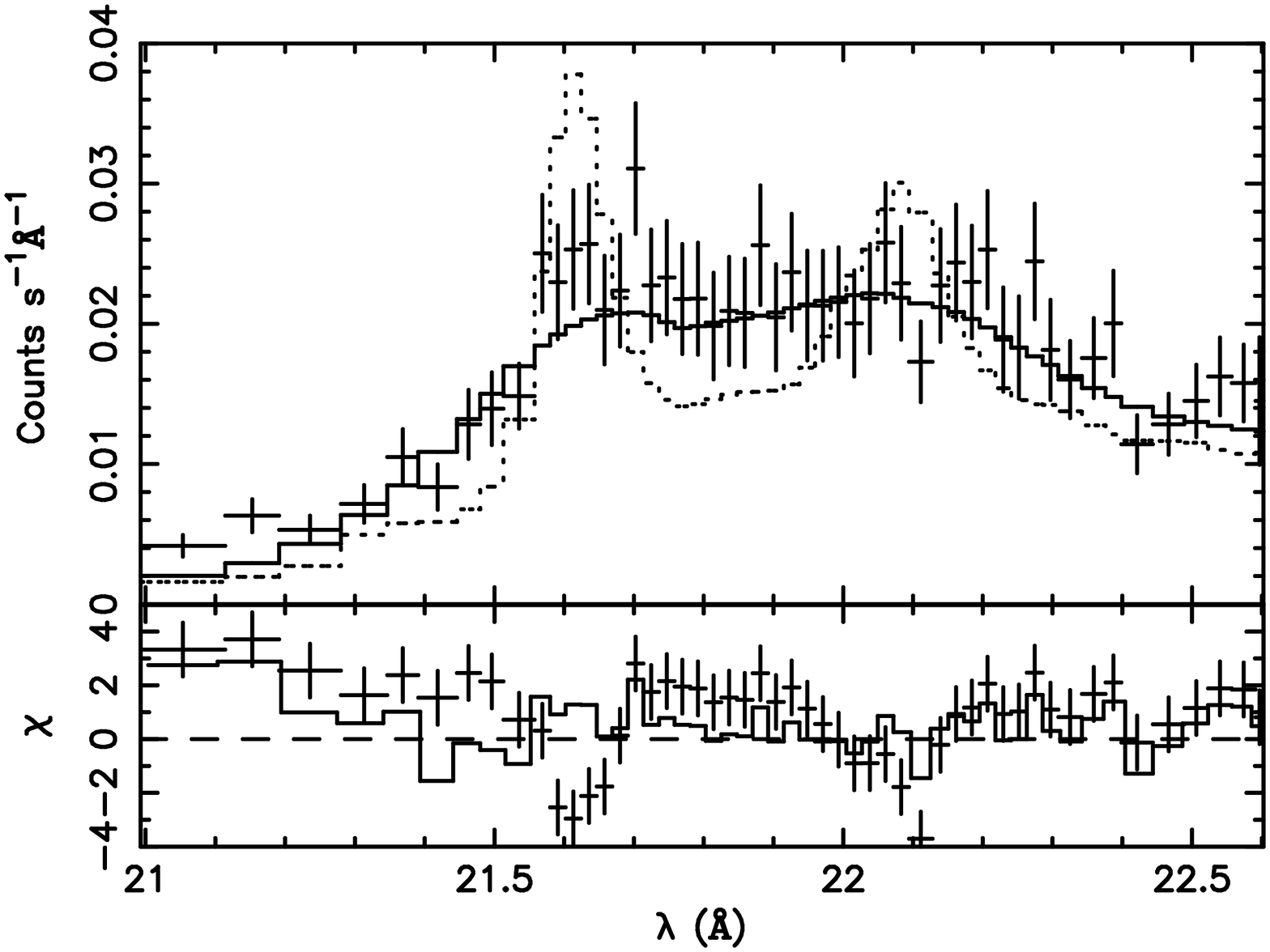,angle=-90,width=8.5cm}
}
\caption{
On the left:
RGS1 spectrum showing O VII and O VIII line emission with line emission 
modelled by broadened gaussian shaped lines. On the right: 
The O VII triplet. The solid line indicate the best fit model with thermal
Doppler broadening. The dotted line shows the best fit model without
broadening. The residuals are shown as a connected line and crosses
respectively. 
\label{fig-rgs}}
\end{figure*}

The RGS1 spectrum (Fig.~\ref{fig-rgs}) shows that the emission
around 0.66~keV is dominated by O~VII He$\beta$\ emission
(O~VIII Ly$\alpha :$\,O~VII He$\beta \simeq 1: 1.6$). 
Ne is detected in the RGS2 spectrum, but Mg and Si are 
too weak for the RGS instruments.
No evidence for Fe~XVII line emission is seen at
15.01~\AA, 16.78~\AA, 17.0~\AA, or 17.10~\AA, presumably because
Fe has not yet reached the Fe~XVII charge state.
The O~VII emission is well described by $n= i \rightarrow 1$,
line ratios that scale with the ratios of the oscillator strengths, as is expected for \Te~$\gtrsim 1$~keV. 

In order to measure the ion temperature through 
the thermal Doppler broadening we fitted the RGS1 
spectrum in the range from $21.0-22.3$~\AA, dominated
by oxygen line emission, with six absorbed gaussians 
and a bremsstrahlung continuum with \Te$=1.5$~keV
fixed to the continuum outside the fitted range
\citep[$N_{\rm H}=6.8\times10^{20}$~cm$^{-2}$, ][]{dubner02}.
The six gaussian components had centroids fixed at the energies of
the bright O~V, O~VI, and O~VII lines. The ratios of those lines
were fixed according to calculations with the FAC atomic code \citep{gu02},
for \Te$=1.5$~keV and a grid of \net\ values between 
$\log (n_{\rm e}t) = 9.0 - 9.4$.
The line broadening was taken to be proportional to the line energies.
Spectral fitting was done with spectral fitting package \xspec\ \citep{xspec}.
This allowed us to use the C-statistic, 
which is the maximum likelihood statistic appropriate for 
Poisson noise \citep{cash79}.

The best fits corresponded to $\log (n_{\rm e}t) = 9.18$
with a maximum likelihood statistic of $C = 93.6$ for
104 data bins, with a possible
range of $\log (n_{\rm e}t) = 9.04 - 9.30$. This is somewhat lower,
but arguably more accurate, than the value derived from the CCD spectra.
It is also consistent with the measured oxygen Ly$\alpha$\ to He$\beta$\
ratio with the RGS. 
For higher value of $\log (n_{\rm e}t)$ the emission would be
much more dominated by the O~VII resonance line at 21.6~\AA, whereas the
RGS1 spectrum indicates a substantial contribution  of forbidden line
emission at 22.1~\AA\ due to O~VI inner shell ionization and excitation.

The line broadening needed to obtain an acceptable fit is 
$\sigma_E = 3.4\pm 0.5$~eV (68\% confidence), 
if fitted within the  wavelength range 21.4-22.6~\AA. 
As Fig.~\ref{fig-rgs} shows the broadening underpredicts the wing of 
the resonance line between 21.0-21.4~\AA, 
which is a possible indication that the oxygen ions may not be completely
thermalized (i.e. are not fully described by a Maxwellian distribution).
The statistical confidence of the detection of line broadening is 
at the $9.8\sigma$\ level ($\Delta C = 96.1$). The 99.73\% (3$\sigma$) 
confidence range is $2.2 - 6.5$~eV. 
Even if we do not fix the line ratios and optimize each individual line,
we obtain $\sigma_E = 2.5\pm 0.4$~eV 
with a statistical confidence still at the  $6.5\sigma$\ level.
% RGSXSRC gives 2.600000E-03
During our analysis we have compared different methods of analyzing the data,
with various wavelength fitting ranges. This has given us some feeling
of the systematic uncertainties involved. We estimate that the
systematic error on $\sigma_E$ is about 0.7~eV.

\section{Discussion}
Our analysis of CCD and grating data of a bright knot in \snks\ obtained by 
\xmm\ 
shows that \Te$\sim1.5$~keV and 
$\log (n_{\rm e}t) = 9.04 - 9.30$
The Doppler broadening at 574~eV is 
$\sigma_E = 3.4\pm0.5$~eV,
which corresponds to an oxygen temperature of $kT_{oxygen} = 530\pm150$~keV
\citep[e.g. eq.~10.69][]{rybicki}.
This indicates a shock velocity of $>4000$~km/s, 
if no significant temperature equilibration has taken place, but allowing
for the possibility of some adiabatic cooling (eq.~\ref{eq-shocks}).
The implied shock velocity is higher than the $\sim 3000$~km/s indicated by
recent H$\alpha$\ measurements \citep{ghavamian02},
but given the systematic errors both measurements agree.

The excess emission in the wings of the O~VII line emission,
although possibly due to calibration uncertainties concerning 
the wings of the intrinsic RGS line profiles, may also be real
and caused by a lack of complete oxygen thermalization.
This is not unexpected as the oxygen 
self equilibration time at high temperatures
is comparable to the oxygen-proton equilibration time.

To conclude, 
the measured high oxygen ion temperature is a clear indication that the shock 
heating processes resulted in only a small degree (5\%) of electron-ion 
equilibration at the shock front, and
that the subsequent equilibration process is slow.
This also has some bearing on the acceleration of
cosmic rays in high Mach number shocks, which are thought to be injected
into the shock acceleration process 
from the pool of the thermal gas behind the shock front.
A low electron-ion equilibration will make it relatively more difficult for
electrons than for ions to be accelerated.
However, the southwestern and northeastern limbs of \snks\ are a 
prime examples of efficient electron acceleration
as the X-ray emission from these parts is dominated by synchrotron emission 
\citep{koyama95}.
Although we could not directly measure the electron temperature at those sites,
the electron temperature close to the northwestern limb seems to be 
similar to that of northeastern knot.
However, the notion of a thermal pool from which particles are accelerated is 
likely to be too simple,
as recent simulations seem to indicate \citep{schmitz02a}.

\acknowledgements
We thank John Peterson for making to us available his
Monte Carlo Code which helped us verify the validity of our results.
JV and MFG were supported for this work by the NASA
through Chandra Postdoctoral Fellowship Award Number PF0-10011 \& PF0-10014
issued by the Chandra X-ray Observatory Center, which is operated by the
Smithsonian Astrophysical Observatory  for NASA under contract
NAS8-39073.
JML was supported by the NASA Contract S 92540F from the XMM GI Program 
and by basic research funds of the Office of Naval Research.

%\bibliographystyle{apj}
%\bibliography{apj-jour,snrs}

\begin{thebibliography}{25}
\expandafter\ifx\csname natexlab\endcsname\relax\def\natexlab#1{#1}\fi

\bibitem[{{Anders} \& {Grevesse}(1989)}]{anders89}
{Anders}, E. \& {Grevesse}, N. 1989, \gca, 53, 197

\bibitem[{{Arnaud}(1996)}]{xspec}
{Arnaud}, K.~A. 1996, in ASP Conf. Ser. 101: Astronomical Data Analysis
  Software and Systems V, Vol.~5, 17

\bibitem[{{Cash}(1979)}]{cash79}
{Cash}, W. 1979, \apj, 228, 939

\bibitem[{{den Herder} {et~al.}(2001)}]{denherder01}
{den Herder}, J.~W. {et~al.} 2001, \aap, 365, L7

\bibitem[{{Dubner} {et~al.}(2002){Dubner}, {Giacani}, {Goss}, {Green}, \&
  {Nyman}}]{dubner02}
{Dubner}, G.~M., {Giacani}, E.~B., {Goss}, W.~M., {Green}, A.~J., \& {Nyman},
  L.-{\AA}. 2002, \aap, 387, 1047

\bibitem[{{Dyer} {et~al.}(2001){Dyer}, {Reynolds}, {Borkowski}, {Allen}, \&
  {Petre}}]{dyer01}
{Dyer}, K.~K., {Reynolds}, S.~P., {Borkowski}, K.~J., {Allen}, G.~E., \&
  {Petre}, R. 2001, \apj, 551, 439

\bibitem[{{Ghavamian} {et~al.}(2001){Ghavamian}, {Raymond}, {Smith}, \&
  {Hartigan}}]{ghavamian01}
{Ghavamian}, P., {Raymond}, J., {Smith}, R.~C., \& {Hartigan}, P. 2001, \apj,
  547, 995

\bibitem[{{Ghavamian} {et~al.}(2002){Ghavamian}, {Winkler}, {Raymond}, \&
  {Long}}]{ghavamian02}
{Ghavamian}, P., {Winkler}, P.~F., {Raymond}, J.~C., \& {Long}, K.~S. 2002,
  \apj, 572, 888

\bibitem[{{Gu}(2002)}]{gu02}
{Gu}, M.~F. 2002, \apjl, 579, L103

\bibitem[{{Hughes} {et~al.}(2000){Hughes}, {Rakowski}, \&
  {Decourchelle}}]{hughes00b}
{Hughes}, J.~P., {Rakowski}, C.~E., \& {Decourchelle}, A. 2000, \apjl, 543, L61

\bibitem[{{Itoh}(1984)}]{itoh84}
{Itoh}, H. 1984, \apj, 285, 601

\bibitem[{{Jansen} {et~al.}(2001)}]{jansen01}
{Jansen}, F. {et~al.} 2001, \aap, 365, L1

\bibitem[{{Kaastra} {et~al.}(1996){Kaastra}, {Mewe}, \&
  {Nieuwenhuijzen}}]{kaastra96}
{Kaastra}, J.~S., {Mewe}, R., \& {Nieuwenhuijzen}, H. 1996, in Proc. of the
  11th Coll. on UV and X-ray, UV and X-ray Spectroscopy of Astrophysical and
  Laboratory Plasmas, ed. K. Yamashita \& T. Watanabe (Tokyo:Universal Academy
  Press), 411

\bibitem[{{Koyama} {et~al.}(1995){Koyama}, {Petre}, {Gotthelf}, {Hwang},
  {Matsuura}, {Ozaki}, \& {Holt}}]{koyama95}
{Koyama}, K., {Petre}, R., {Gotthelf}, E.~V., {Hwang}, U., {Matsuura}, M.,
  {Ozaki}, M., \& {Holt}, S.~S. 1995, \nat, 378, 255

\bibitem[{{Laming}(2001)}]{laming01b}
{Laming}, J.~M. 2001, \apj, 563, 828

\bibitem[{{Laming} {et~al.}(1996){Laming}, {Raymond}, {McLaughlin}, \&
  {Blair}}]{laming96}
{Laming}, J.~M., {Raymond}, J.~C., {McLaughlin}, B.~M., \& {Blair}, W.~P. 1996,
  \apj, 472, 267

\bibitem[{{Long} {et~al.}(2003){Long}, {Reynolds}, {Raymond}, {Winkler},
  {Dyer}, \& {Petre}}]{long03}
{Long}, K.~S., {Reynolds}, S.~P., {Raymond}, J.~C., {Winkler}, P.~F., {Dyer},
  K.~K., \& {Petre}, R. 2003, \apj, in press

\bibitem[{{McKee} \& {Hollenbach}(1980)}]{mckee80}
{McKee}, C.~F. \& {Hollenbach}, D.~J. 1980, \araa, 18, 219

\bibitem[{{Michael} {et~al.}(2002){Michael}, {Zhekov}, {McCray}, {Hwang},
  {Burrows}, {Park}, {Garmire}, {Holt}, \& {Hasinger}}]{michael02}
{Michael}, E., {Zhekov}, S., {McCray}, R., {Hwang}, U., {Burrows}, D.~N.,
  {Park}, S., {Garmire}, G.~P., {Holt}, S.~S., \& {Hasinger}, G. 2002, \apj,
  574, 166

\bibitem[{{Raymond} {et~al.}(1995){Raymond}, {Blair}, \& {Long}}]{raymond95}
{Raymond}, J.~C., {Blair}, W.~P., \& {Long}, K.~S. 1995, \apjl, 454, L31

\bibitem[{{Rybicki} \& {Lightman}(1979)}]{rybicki}
{Rybicki}, G.~B. \& {Lightman}, A.~P. 1979, {Radiative processes in
  astrophysics} (New York, Wiley-Interscience, 1979.~393 p.)

\bibitem[{{Schmitz} {et~al.}(2002){Schmitz}, {Chapman}, \&
  {Dendy}}]{schmitz02a}
{Schmitz}, H., {Chapman}, S.~C., \& {Dendy}, R.~O. 2002, \apj, 570, 637

\bibitem[{{Truelove} \& {McKee}(1999)}]{truelove99}
{Truelove}, J.~K. \& {McKee}, C.~F. 1999, \apjs, 120, 299

\bibitem[{{Vink} {et~al.}(2000){Vink}, {Kaastra}, {Bleeker}, \&
  {Preite-Martinez}}]{vink00a}
{Vink}, J., {Kaastra}, J.~S., {Bleeker}, J.~A.~M., \& {Preite-Martinez}, A.
  2000, \aap, 354, 931

\bibitem[{{Winkler} {et~al.}(2003){Winkler}, {Gupta}, \& {Long}}]{winkler02}
{Winkler}, P.~F., {Gupta}, G., \& {Long}, K.~S. 2003, \apj, submitted,
  (astroph/0208415)

\end{thebibliography}
%\clearpage

%\input sn1006_xmm_table1a

\end{document}